\documentstyle[times,pramana,epsf,floats]{ias}
\begin{document}
\mark{{Probabilistic Interpretation of Resonant States}{N Hatano, T Kawamoto and J Feinberg}}
\title{Probabilistic Interpretation of Resonant States}

\author{Naomichi Hatano$^1$, Tatsuro Kawamoto$^2$ and Joshua Feinberg$^{3,4}$}
\address{$^1$Institute of Industrial Science, University of Tokyo, Komaba, Meguro, Tokyo 153-8505, Japan\\
$^2$Department of Physics, University of Tokyo, Komaba, Meguro, Tokyo 153-8505, Japan\\
$^3$Department of Physics, University of Haifa at Oranim, Qiryat Tivon 36006, Israel\\
$^4$Department of Physics, Technion, Haifa 32000, Israel}
\keywords{Resonance, Probabilistic Interpretation, Normalization}
\pacs{03.65.Nk}
\abstract{We provide probabilistic interpretation of resonant states.
This we do by showing that the integral of the modulus square of resonance wave functions (\textit{i.e.}, the conventional norm) over a properly expanding spatial domain is independent of time, and therefore leads to probability conservation. 
This is in contrast with the conventional employment of a bi-orthogonal basis that precludes probabilistic interpretation, since wave functions of resonant states diverge exponentially in space.
On the other hand, resonant states decay exponentially in time, because momentum leaks out of the central scattering area.
This momentum leakage is also the reason for the spatial exponential divergence of resonant state.
It is by combining the opposite temporal and spatial behaviors of resonant states that we arrive at our probabilistic interpretation of these states.
The physical need to normalize resonant wave functions over an expanding spatial domain arises because particles leak out of the region which contains the potential range and escape to infinity, and one has to include them in the total count of particle number.}

\maketitle
\section{Introduction}

The present paper reviews the basic definition of the resonant state in quantum mechanics~\cite{Razavy03} as well as a recent progress on physical interpretation of the resonant state~\cite{Hatano08}.
Presenting partly new calculations, we stress that the particle number, after its proper definition, is conserved for the resonant state.
Hence, the probabilistic interpretation of the square modulus of the wave function remains intact even for the resonant state.

The phenomenon of resonance has been of great importance in elementary-particle physics and nuclear physics~\cite{Gamow28,Siegert39,Peierls59,leCouteur60,Zeldovich60,Humblet61,Rosenfeld61,Humblet62,Humblet64-1,Jeukenne64,Humblet64-2,Mahaux65,Rosenfeld65,Wigner55,Narnhofer81,Amrein87,Carvalho02,Ahmed04,Kelkar04,Jain05,Amrein06,Landau77,Newton82,Brandas89,Kukulin89,Kato06}.
In recent years, unstable nuclei, which are nothing but resonant states, are experimentally sought after; see \textit{e.g.}, Refs.~\cite{Morita04,Morita07}.
It has been also realized in condensed-matter physics that the resonance reveals itself in electronic conduction in mesoscopic systems~\cite{Kobayashi02,Kobayashi03,Kobayashi04,Sato05,Nishino07,Nishino09}.
These facts, among others, have revived theoretical interest in the phenomenon of resonance in quantum mechanics~\cite{Nakamura07}.

The wave function of a resonant state has been discussed greatly since the early stage of quantum mechanics.
As will be shown at the end of Sec.~3 below, it exponentially diverges in the distance from the scattering center.
Although this is an inevitable conclusion, the divergence has hindered the probabilistic interpretation of the square modulus of the wave function.
A bi-orthonormal space of the left- and right-eigenfunction was thereby introduced to define a convergent norm~\cite{Zeldovich60,Hokkyo65,Romo68,Berggren70,Gyarmati71,Romo80,Berggren82,Berggren96,Madrid05}.

The main purpose of the present paper is to stress that the probabilistic interpretation of the square modulus of the wave function is nonetheless possible~\cite{Hatano08}.
We show that particles leak from the central region in the resonant state.
The exponential divergence in space of the wave function is a direct consequence of the leaking particles.
The particle number (in the form of the square modulus of the wave function) is conserved when we take the leak into account.
More specifically, we expand the region where we count the particle number, with the speed of the leaking particles.
For arbitrary superpositions of states, we introduce a definition of the speed of the leaking particles.
With this proper definition of the speed, we show that the integral of the square modulus of the wave function over the expanding region is independent of time.

The paper is organized as follows.
In Sec.~2, we review the definition of the resonant state as an eigenstate of the Schr\"{o}dinger equation.
We here introduce the Siegert condition that there exist only out-going waves away from the scattering region.
We then present in Sec.~3 a physical view of the resonant state with the Siegert condition.
The resonant state will be described as a state that decays in time exponentially with momentum leaks from the scattering region.
On the basis of this view, we show in Sec.~4 that the particle number defined as the square modulus of the wave function is indeed conserved.
The probabilistic interpretation of the square modulus of the wave function is thus intact even for the resonant state.
The last section is devoted to a summary.

\section{Direct definition of quantum resonance}
\label{sec2}

In many textbooks, resonance is defined as a pole of the $S$ matrix.
It is, however, more convenient to define a resonant state as an eigenstate of the Schr\"{o}dinger equation with the Siegert condition given below.
Let us briefly review the connection of the two definitions in the present section.

Consider, for simplicity, a one-dimensional system with a scattering potential localized around the origin.
An incident wave $Ae^{ikx}$ from the left ($\mathop{\mathrm{Re}}k>0$) will be scattered to result in a reflection wave $Be^{-ikx}$ and a transmission wave $Ce^{ikx}$.
When we properly connect these three waves at the scattering potential, the coefficients $A$, $B$ and $C$ become dependent on the wave number $k$, or on the energy $E$ through the dispersion relation $E(k)$ of a free space (specifically $E(k)=\hbar^2k^2/(2m)$ in the following example).

Elements of the $S$ matrix are then given by the ratios $B/A$ and $C/A$.
The resonance is defined as a pole of the $S$ matrix in the complex wave number plane or in the complex energy plane.
In fact, zeros of $A$ give poles of the $S$ matrix.

\begin{figure}
\epsfxsize=0.6\textwidth
\centerline{\epsfbox{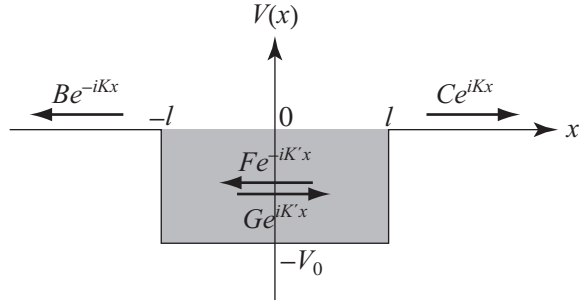}}
\vspace*{\baselineskip}
\caption{A square-well potential with the Siegert condition of out-going waves only.}
\label{fig1}
\end{figure}
The above two-step procedure of (i) solving a scattering problem for real $k$ and $E$ and (ii) finding zeros of $A$ in the complex $k$ and $E$ plane, can be reduced to just one step of solving a scattering problem for complex $k$ and $E$ with the condition $A\equiv0$ required from the very beginning.
The condition $A\equiv0$ means that we have out-going waves only.
This boundary condition is referred to as the Siegert condition.
The above argument shows that a pole of the $S$ matrix is indeed given by a state with the Siegert boundary condition of out-going waves only~\cite{Siegert39,Peierls59}.

For some readers who might wonder if a solution with the Siegert condition is possible, let us solve a simple problem shown in Fig.~\ref{fig1}.
The Hamiltonian is given by
\begin{equation}\label{eq110}
{\mathcal{H}}=-\frac{\hbar^2}{2m}\frac{d^2}{dx^2}+V(x)
\end{equation}
with the square-well potential
\begin{equation}\label{eq10}
V(x)=\left\{\begin{array}{ll}
-V_0<0 & \quad \mbox{for $|x|<l$,} \\
0 & \quad \mbox{otherwise.}
\end{array}\right.
\end{equation}
We assume the form of the solution as
\begin{equation}\label{eq20}
\Phi(x)=\left\{\begin{array}{ll}
Be^{-iKx} & \quad \mbox{for $x\leq -l$,}\\
Fe^{-iK'x}+Ge^{iK'x}& \quad\mbox{for $|x|<l$,} \\
Ce^{iKx} & \quad \mbox{for $x\geq l$,}
\end{array}\right.
\end{equation}
where $K$ and $K'$ are wave numbers in the respective region, related to each other as
\begin{equation}\label{eq25}
{K'}^2-\frac{2m}{\hbar^2}V_0=K^2.
\end{equation}
We here used uppercase letters for the wave numbers in order to emphasize that they are complex numbers.

\begin{figure}
\epsfxsize=\textwidth
\centerline{\epsfbox{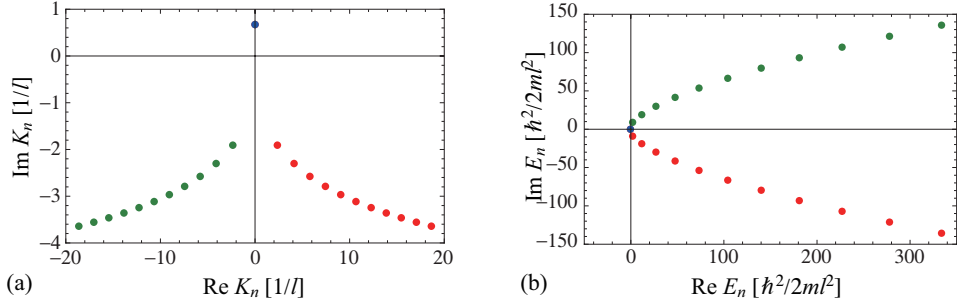}}
\vspace*{\baselineskip}
\caption{Solutions $K_n$ and $E_n$ for the square-well potential with the Siegert condition.
We set $V_0=1[\hbar^2/2ml^2]$.
The dot on the positive imaginary $K$ axis and on the negative real $E$ axis is a bound state.
(It has a small real negative value of the energy.)
The dots on the fourth quadrant of the $K$ plane and on the lower half of (the second Riemann sheet of) the $E$ plane are resonant states, whereas the dots on the third quadrant of the $K$ plane and on the upper half of (the second Riemann sheet of) the $E$ plane are anti-resonant states.}
\label{fig2}
\end{figure}
Since the potential is an even function, we can classify resonant solutions into odd and even ones.
We set $B=-C$ and $F=-G$ for odd solutions and $B=C$ and $F=G$ for even solutions.
Upon solving the Schr\"{o}dinger equation, we require the usual conditions that $\Phi(x)$ and $\Phi'(x)$ are continuous at $x=l$.
First, the continuity of $\Phi(x)$ at $x=l$ gives odd solutions of the form
\begin{equation}\label{eq27}
\Phi(x)=2iG\times\left\{\begin{array}{ll}
\sin (K'x)&\quad\mbox{for $|x|<l$}, \\
\mathop{\mathrm{sgn}}x\sin (K'l) e^{iK(|x|-l)}&\quad\mbox{for $|x|>l$},
\end{array}\right.
\end{equation}
and even solutions of the form
\begin{equation}\label{eq28}
\Phi(x)=2G\times\left\{\begin{array}{ll}
\cos (K'x)&\quad\mbox{for $|x|<l$}, \\
\cos (K'l) e^{iK(|x|-l)}&\quad\mbox{for $|x|>l$}.
\end{array}\right.
\end{equation}
Then, the continuity of $\Psi'(x)$ at $|x|=l$ gives the equations
\begin{eqnarray}\label{eq30}
K&=&-iK'\cot(K'l)\quad\mbox{for odd solutions,}
\\\label{eq31}
K&=&iK'\tan(K'l)\quad\;\;\mbox{for even solutions,}
\end{eqnarray}
which should be solved together with Eq.~(\ref{eq25}).
We solved them numerically to obtain $K_n$ and $E_n$ plotted in Fig.~\ref{fig2}.
First of all, the wave function of the form~(\ref{eq20}) represent a bound state when $K$ is a positive pure imaginary.
We then have pairs of solutions on the lower half of the $K$ plane.
One of each pair on the fourth quadrant is called a resonant state, whereas the other one on the third quadrant is called an anti-resonant state.
(In the context of electronic conduction of mesoscopic systems, some call a dip in the conductance an anti-resonance.
This is completely unrelated to the anti-resonant state in the present paper.)

\begin{figure}
\epsfxsize=0.8\textwidth
\centerline{\epsfbox{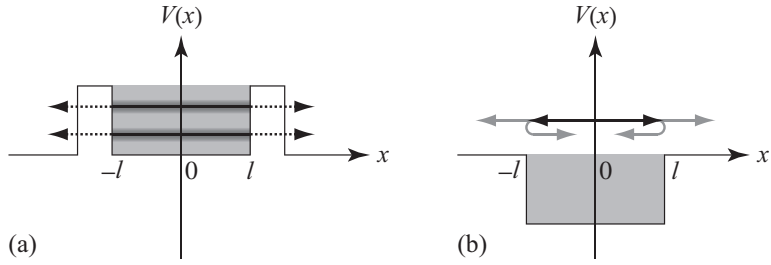}}
\vspace*{\baselineskip}
\caption{(a) A particle trapped in the shaded area can escape the trap because of tunneling effect.
The coupling between the bound state inside the trap and the continuum outside makes the bound state unstable and results in a resonance width.
(b) A particle with a positive kinetic energy is partially reflected at the potential edges $x=\pm l$ and hence can be ''trapped" in the potential well for a while.}
\label{fig3}
\end{figure}

Some readers might wonder why the present potential without any potential barriers can support resonances.
Indeed, a conventional view of quantum-mechanical resonance might be as follows for a potential with some barriers such as exemplified in Fig.~\ref{fig3}(a);
a particle that would be completely trapped classical-mechanically in the range $|x|<l$ and hence would be a bound state quantum-mechanically, can in fact escape the trap because of tunneling effect, and this coupling to the continuum outside the barriers causes a resonance width to the would-be bound state; see Ref.~\cite{Hatano08} for such an example.
Such a resonance shows up in the Hermitian problem as narrow transmission resonances at energies below the barrier, where the transmission coefficient shoots up to unity. 

Nevertheless, the potential well in Fig.~\ref{fig3}(b) can support resonances that we discuss here and hereafter.
The square-well potential can partially trap a quantum-mechanical particle because the wave of a particle that departs the origin $x=0$ towards outside is partially reflected at the potential edges $x=\pm l$ even when its kinetic energy is positive.
This reflection at the potential edges plays a role of an effective potential barrier.
Indeed, this effect would cause the Anderson localization if we had infinitely many potential wells~\cite{Ziman79}.
We will argue in the next section that the complex eigen-wave-number and eigenenergy of a resonant state are directly linked to out-going momentum fluxes from the central range of the scattering potential.
This view holds for potential of both types in Figs.~\ref{fig3}(a) and~(b).
The transmission coefficient associated with the square-well potential in Fig.~\ref{fig3}(b) is very close to unity for any positive energies with broad local maxima around where it reaches unity. 
These are not well-defined transmission resonances in the conventional sense.
Nonetheless, they are in fact the Hermitian shadows of the resonances that we discuss in the present paper.
Interested readers might want to follow the discussion hereafter with their own example potential of the type in Fig.~\ref{fig3}(a);
it will not cause much difference.

At the end of the present section, we stress that the above definition on the basis of the Siegert condition gives the resonant (and anti-resonant) states as eigensolutions of the Schr\"{o}dinger equation.
The wave function of each resonant or anti-resonant state is explicitly given by substituting the corresponding solution in the $K$ plane in Eq.~(\ref{eq27}) or~(\ref{eq28}).

\section{Physical view of quantum resonance}
\label{sec3}

In the present section, we give a physical view of the resonant and anti-resonant states as defined in the previous section.
We begin with discussing the non-Hermiticity of the Hamiltonian of open quantum systems.
For simplicity again, we consider a one-dimensional system of the form~(\ref{eq110}), where $V(x)$ is a real function with a compact support around the origin.

We first define the expectation value of the Hamiltonian with respect to an arbitrary function $\Psi(x)$ in the form
\begin{equation}\label{eq120}
\langle\Psi|{\mathcal{H}}|\Psi\rangle_\Omega
=\int_{-L}^L\Psi(x)^\ast{\mathcal{H}}\Psi(x)dx,
\end{equation}
where $\Omega\equiv[-L,L]$ is a large region whose edges $x=\pm L$ are far away from the support of the scattering potential $V(x)$;
that is, $V(L)=V(-L)=0$.
We may take the limit $L\to\infty$ in the end if we can, but for the moment we keep $L$ to be finite.
(Note here and hereafter that the bra vector is given by the standard definition of being complex conjugate, not given in the dual space spanned by the left eigenvectors~\cite{Zeldovich60,Hokkyo65,Romo68,Berggren70,Gyarmati71,Romo80,Berggren82,Berggren96,Madrid05}.)

We are now interested in the non-Hermiticity of the Hamiltonian;
that is, we would like to know whether $\langle\Psi|{\mathcal{H}}|\Psi\rangle_\Omega$ is real or not.
Clearly, $\langle\Psi|V(x)|\Psi\rangle_\Omega$ is real.
The remaining problem is the reality of the quantity $\langle\Psi|(d^2/dx^2)|\Psi\rangle_\Omega$.
For this purpose, we carry out a partial integration:
\begin{eqnarray}\label{eq130}
\left\langle\Psi\left|\frac{d^2}{dx^2}\right|\Psi\right\rangle
&=&
\int_{-L}^L\Psi(x)^\ast\Psi''(x)dx
\nonumber\\
&=&
\left[\Psi(x)^\ast\Psi'(x)\right]_{x=-L}^L
-\int_{-L}^L|\Psi'(x)|^2dx.
\end{eqnarray}
The second term on the right-hand side is obviously real.
On the other hand, we rewrite the first term as follows;
\begin{equation}\label{eq135}
\left[\Psi(x)^\ast\Psi'(x)\right]_{x=-L}^L
=\frac{i}{\hbar}\left(\left.\Psi(x)^\ast\hat{p}\Psi(x)\right|_{x=L}-\left.\Psi(x)^\ast\hat{p}\Psi(x)\right|_{x=-L}\right),
\end{equation}
where $\hat{p}=(\hbar/i)d/dx$ is the momentum operator.
Therefore, we have the identity
\begin{equation}\label{eq140}
\mathop{\mathrm{Im}}\langle\Psi|{\mathcal{H}}|\Psi\rangle_\Omega
=-\frac{\hbar}{2m}\mathop{\mathrm{Re}}\langle\Psi|\hat{p}_{\mathrm{n}}|\Psi\rangle_{\partial\Omega},
\end{equation}
where we used the bra-ket notation, which holds for any space dimension $d$.
Here, $\hat{p}_{\mathrm{n}}$ is the normal component of the momentum on the boundary of $\Omega$;
in the present case, $\hat{p}_{\mathrm{n}}=\hat{p}$ at $x=L$ and $\hat{p}_{\mathrm{n}}=-\hat{p}$ at $x=-L$.
We also let $\langle\Psi|\cdot|\Psi\rangle_{\partial\Omega}$ denote the expectation value on the boundary of $\Omega$.
We derived the identity~(\ref{eq140}) in one dimension but it holds in arbitrary dimensions, in fact.
(When we put $\Psi$ to each resonant state $\Phi_n$, the identity~(\ref{eq140}) reduces to the one derived previously~\cite{Berggren87,Moiseyev90,Masui99}.)

The equality~(\ref{eq140}) shows that the non-Hermiticity of the Hamiltonian comes from the momentum leak at the boundary of the region.
Its meaning becomes clearer when we extend the above argument to solutions of the time-dependent Schr\"{o}dinger equation, $\Psi(x,t)$.
For a solution of the equation
\begin{equation}\label{eq150}
i\hbar\frac{\partial}{\partial t}\Psi(x,t)={\mathcal{H}}\Psi(x,t),
\end{equation}
we can easily prove the identity~\cite{Bohm89}
\begin{equation}\label{eq160}
\frac{d}{dt}\left\langle\Psi|\Psi\right\rangle_\Omega=\frac{2}{\hbar}\mathop{\mathrm{Im}}\langle\Psi|{\mathcal{H}}|\Psi\rangle_\Omega,
\end{equation}
and thus we have 
\begin{equation}\label{eq170}
\frac{d}{dt}\left\langle\Psi|\Psi\right\rangle_\Omega=-\frac{1}{m}\mathop{\mathrm{Re}}\langle\Psi|\hat{p}_{\mathrm{n}}|\Psi\rangle_{\partial\Omega}.
\end{equation}
This has a very plausible interpretation;
the decreasing rate (for $\mathop{\mathrm{Re}}\langle\Psi|\hat{p}_{\mathrm{n}}|\Psi\rangle_{\partial\Omega}>0$) of the particle numbers in the region $\Omega$ is proportional to the momentum leak at the boundary of the region, $\partial\Omega$.
Equations~(\ref{eq140}) and~(\ref{eq160}) show that the two quantities are indeed linked to the non-Hermiticity of the system.
This is how an open quantum system becomes non-Hermitian.

The above interpretation leads to the following physical view of the resonant and anti-resonant states.
In a resonant state, there is a leak on the boundary of the region $\Omega$ ($\mathop{\mathrm{Re}}K_n>0$) and hence the particle number in $\Omega$ decays in time ($\mathop{\mathrm{Im}}E_n<0$).
In an anti-resonant state, on the other hand, particles come into the region $\Omega$ ($\mathop{\mathrm{Re}}K_n<0$) and hence the particle number in $\Omega$ grows in time ($\mathop{\mathrm{Im}}E_n>0$).
This view of the resonant and anti-resonant states applies both to a trapping potential in Fig.~\ref{fig3}(a) and to a potential in Fig.~\ref{fig3}(b).

Incidentally, the dispersion relation $E_n=(\hbar K_n)^2/2m$ is followed by
\begin{equation}\label{eq180}
\mathop{\mathrm{Im}}E_n= \frac{\hbar^2}{m}\mathop{\mathrm{Re}}K_n\mathop{\mathrm{Im}}K_n.
\end{equation}
The argument in the previous paragraph then gives $\mathop{\mathrm{Im}}K_n<0$ for all resonant and anti-resonant states.
This, however, means that the wave function~(\ref{eq20}) diverges in the form
\begin{equation}\label{eq190}
\left|\Phi_n(x)\right|^2\sim e^{2|\mathop{\mathrm{Im}}K_n||x|}\quad\mbox{as $|x|\to\infty$.}
\end{equation}
This divergence indeed has hindered the probabilistic interpretation of the wave function of the resonant state.
Nevertheless, we show in the next section that the probabilistic interpretation is actually possible.

\section{Particle number conservation}
\label{sec4}

\begin{figure}
\epsfxsize=0.9\textwidth
\centerline{\epsfbox{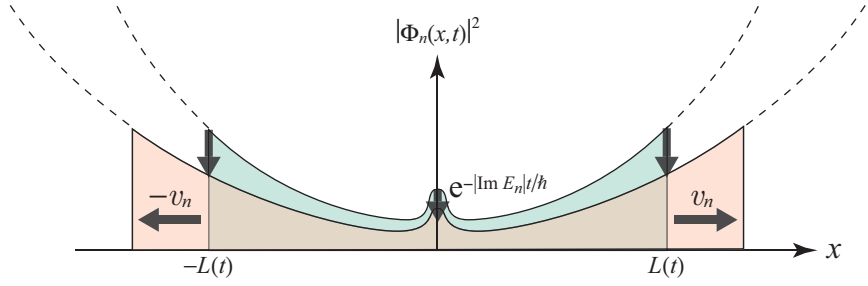}}
\vspace*{\baselineskip}
\caption{A schematic view of the particle number conservation of a resonant state.
The particle number decays in time as $\exp(-\mathop{\mathrm{Im}}E_nt/\hbar)$ over the entire space, but it is conserved when we expand the integration region with the speed $v_n=\mathop{\mathrm{Re}}K_n\hbar/m$.}
\label{fig4}
\end{figure}

In the present section, we show that the particle number is conserved for the resonant and anti-resonant states.
For simplicity again, we proceed in one dimension.
The basic idea is as follows (Fig.~\ref{fig4}).
Consider a resonant eigenstate
\begin{equation}
\Phi_n(x,t)=\Phi_n(x)e^{-iE_nt/\hbar}
\end{equation}
with the eigen-wave-number $K_n$ and the eigenenergy $E_n$.
At one point of time, which we refer to as $t=0$, we count the particle number of the state over a finite region of space, $\Omega=[-L(0),L(0)]$.
In order to compare the particle number at a later time, we have to take into account the fact that the particles are fleeing from the region in a resonant state, as we pointed out at the end of the previous section.
The fleeing speed outside the potential range may be considered as
\begin{equation}\label{eq210}
v_n\equiv\frac{\hbar}{m}\mathop{\mathrm{Re}}K_n.
\end{equation}
Therefore, we should expand the region $\Omega(t)=[-L(t),L(t)]$ for counting the particle number.
The speed of the expansion should be equal to $v_n$; that is,
\begin{equation}\label{eq215}
L(t)\equiv v_nt + L(0).
\end{equation}

We show below that the particle number in the expanding region is indeed constant~\cite{Hatano08}.
Let us calculate the time derivative of the particle number in the expanding region:
\begin{eqnarray}\label{eq220}
\frac{d}{dt}\left\langle\Psi|\Psi\right\rangle_{\Omega(t)}
&=&
\frac{d}{dt}\int^{L(t)}_{-L(t)}\left|\Psi(x,t)\right|^2dx
\nonumber\\
&=&-\frac{1}{m}\mathop{\mathrm{Re}}\langle\Psi|\hat{p}_{\mathrm{n}}|\Psi\rangle_{\partial\Omega(t)}
+2\dot{L}(t)\left|\Psi(L(t),t)\right|^2
\nonumber\\
&=&-\frac{2}{m}\mathop{\mathrm{Re}}\langle\Psi|\hat{p}|\Psi\rangle_{x=L(t)}
+2\dot{L}(t)\langle \Psi|\Psi\rangle_{x=L(t)}.
\end{eqnarray}
(Note that we do not take space integration in both terms of the last line.
In higher dimensions, we would need surface integration over the boundary of the region $\Omega$.) 
The first term on the right-hand side of Eq.~(\ref{eq220}) is given by Eq.~(\ref{eq170}) and is equal to the decay rate of the particle number in the region $\Omega$, while the second term comes from the derivative of the integration bounds $\pm L(t)$ and is equal to the increase of the particle number due to the expanding integration region.

Using the explicit form of the wave function~(\ref{eq20}), or
\begin{equation}\label{eq230}
\Psi(x,t)=\Phi_n(x,t)=\left\{\begin{array}{l}
e^{i(K_nx-E_nt/\hbar)}\\
\qquad\qquad\mbox{on the right of the potential range,} \\
\pm e^{i(-K_nx-E_nt/\hbar)}\\
\qquad\qquad\mbox{on the left of the potential range,}
\end{array}\right.
\end{equation}
where the sign of the second line on the right-hand side depends on whether the solution is odd or even, we obtain
\begin{equation}\label{eq240}
\frac{d}{dt}\left\langle\Phi_n|\Phi_n\right\rangle_{\Omega(t)}
=
2\left(-\frac{\hbar}{m}\mathop{\mathrm{Re}}K_n +\dot{L}(t)\right)e^{\mathop{\mathrm{Im}}(K_nL(t)-E_nt/\hbar)}
=0.
\end{equation}
Not only the coefficient but also the time dependent part of the exponent on the right-hand side vanishes because of the dispersion relation $E_n=(\hbar K_n)^2/(2m)$.
Thus we proved that each of the resonant and anti-resonant states conserves the particle number.
In other words, the probabilistic interpretation of the square modulus of the wave function is intact for the resonant and anti-resonant states.

The probabilistic interpretation of pure resonant states should also hold for superpositions of resonant and anti-resonant states.
Thus, let $\Psi(x,t)$ be a generic solution of the time-dependent Schr\"{o}dinger equation~(\ref{eq150}).
Since the identity~(\ref{eq220}) holds for arbitrary such functions $\Psi$, we can make $(d/dt)\left\langle\Psi|\Psi\right\rangle_{\Omega(t)}$ vanish by requiring $L(t)$ to satisfy
\begin{equation}\label{eq250}
\dot{L}(t)=\bar{v}(L(t))
\end{equation}
with the speed of the fleeing particles defined by
\begin{equation}\label{eq260}
\bar{v}(x) \equiv \frac{1}{m}
\frac{\displaystyle \mathop{\mathrm{Re}}\langle\Psi|\hat{p}|\Psi\rangle_{x}}{\displaystyle \langle \Psi|\Psi\rangle_{x}},
\end{equation}
which generalizes Eq.~(\ref{eq210}).
Equation~(\ref{eq250}) constitutes a differential equation for $L(t)$.
Thus, given $\Psi(x,t)$, we can determine $L(t)$ by solving Eq.~(\ref{eq250}) instead of using Eq.~(\ref{eq215}).

As a simple application of Eq.~(\ref{eq260}), beyond the pure resonant case, consider a superposition of two resonant states
\begin{equation}\label{eq270}
\Psi(x,t)=a_1 \Phi_1(x,t)+a_2 \Phi_2(x,t)
\end{equation}
with time-independent coefficients $a_1$ and $a_2$.
Outside the potential range, we have the form~(\ref{eq230})
for each of $\Phi_1$ and $\Phi_2$ with $\mathop{\mathrm{Re}}K_n>0$, $\mathop{\mathrm{Im}}K_n<0$ and $\mathop{\mathrm{Im}}E_n<0$.
Substituting of Eq.~(\ref{eq270}) in Eq.~(\ref{eq260}) then yields, after a straightforward (but somewhat tedious) computation, to the differential equation
\begin{eqnarray}\label{eq290}
\dot{L}&=&\frac{\hbar/m}{\displaystyle
|a_1|^2e^{\mathop{\mathrm{Im}}\Delta(t)}
+|a_2|^2e^{-\mathop{\mathrm{Im}}\Delta(t)}
+2|a_1a_2|\cos\Theta(t)
}
\nonumber\\
&&\times\left\{|a_1|^2e^{\mathop{\mathrm{Im}}\Delta(t)}\mathop{\mathrm{Re}}K_1
+|a_2|^2e^{-\mathop{\mathrm{Im}}\Delta(t)}\mathop{\mathrm{Re}}K_2
\right.
\nonumber\\
&&\phantom{\times}+|a_1a_2|\left[
\mathop{\mathrm{Re}}(K_1+K_2)\cos\Theta(t)
+\mathop{\mathrm{Im}}(K_1-K_2)\sin\Theta(t)\right]\Bigr\},
\end{eqnarray}
where
\begin{eqnarray}\label{eq300}
\Delta(t)&\equiv&\frac{E_1-E_2}{\hbar}t-(K_1-K_2)L(t),
\\
\Theta(t)&\equiv&\mathop{\mathrm{Re}}\Delta(t)-(\arg a_1-\arg a_2).
\end{eqnarray}

\begin{figure}
\hspace*{0.01\textwidth}
\epsfysize=0.44\textwidth
\epsfbox{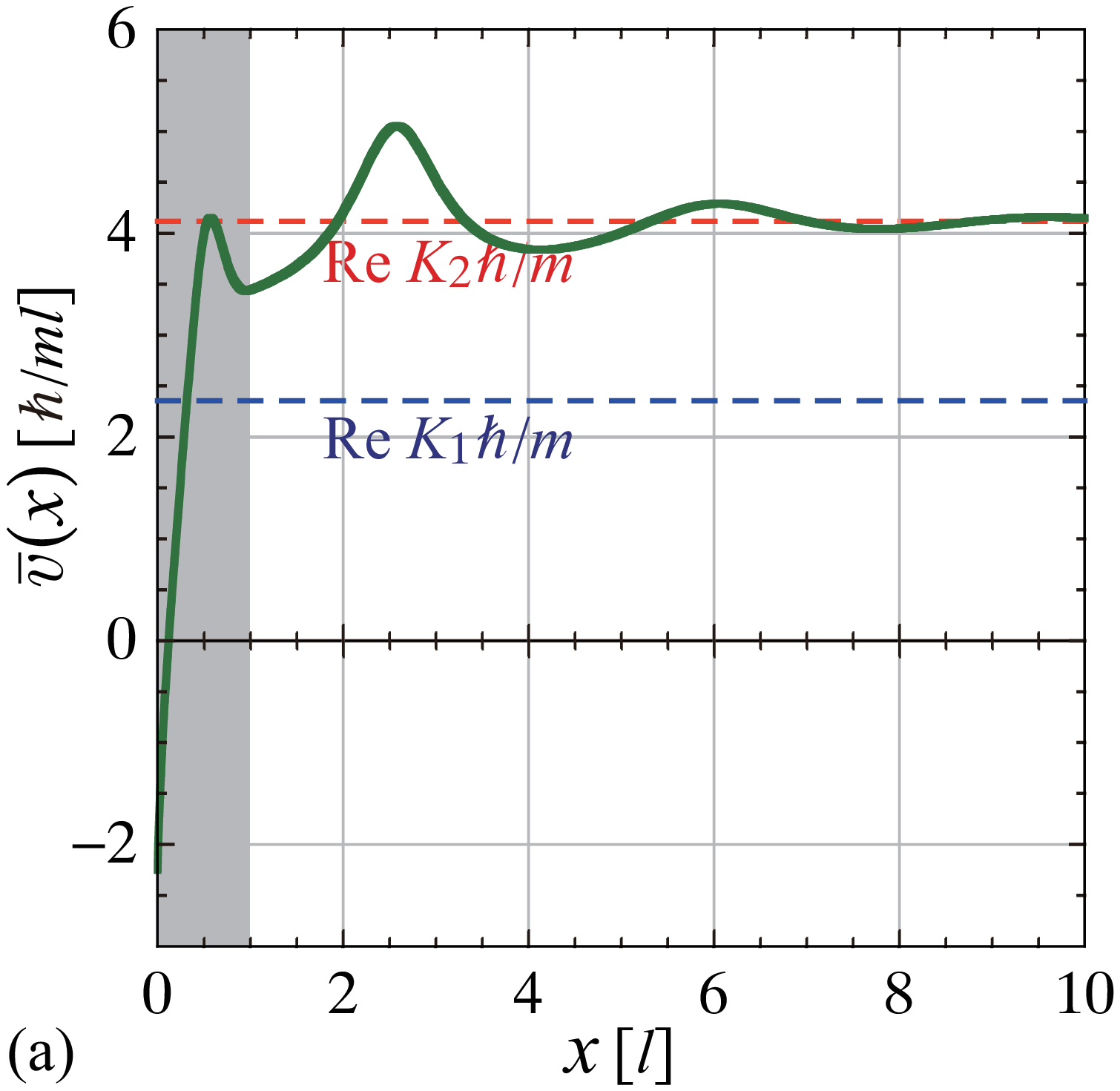}
\hspace*{0.02\textwidth}
\epsfysize=0.44\textwidth
\epsfbox{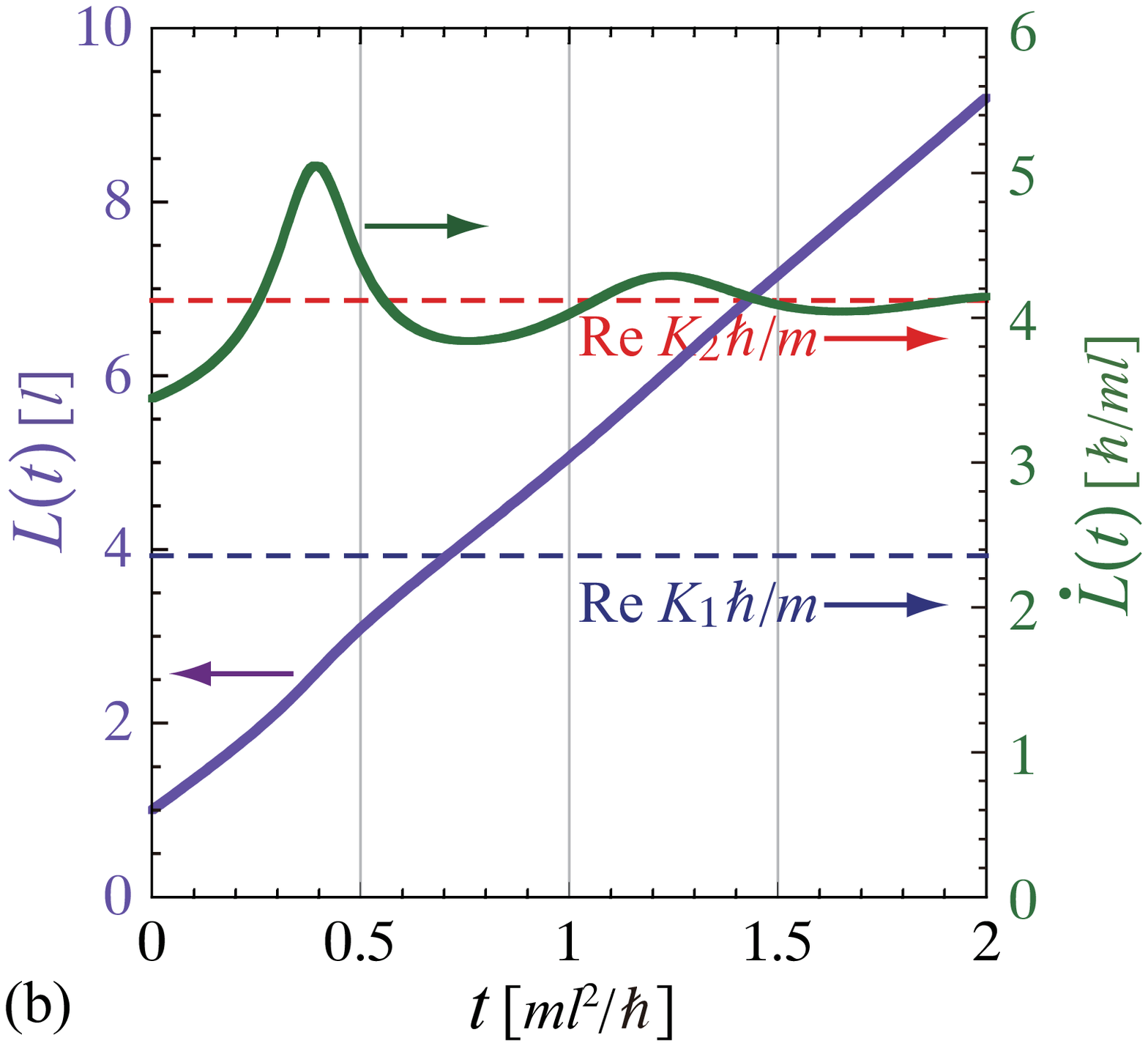}
\vspace*{\baselineskip}
\caption{(a) Enumeration of the fleeing speed~(\ref{eq260}). The shaded area indicates the potential well.
(b) The numerical solution of the differential equation~(\ref{eq290}) with the initial condition $L(0)=l$.}
\label{fig5}
\end{figure}
Let us solve the differential equation~(\ref{eq290}) for the first and second resonant states of the symmetric square well potential~(\ref{eq10}).
The states are depicted as the first and second dots in the fourth quadrant of Fig.~\ref{fig2}(a).
Specifically, we have
\begin{eqnarray}
&&K_1\simeq2.356987-i1.909078[1/l]
\nonumber\\
&&\qquad\qquad\qquad\mbox{with}\quad
E_1\simeq1.910812-i8.999349[\hbar^2/2ml^2],
\\
&&K_2\simeq4.119962-i2.301222[1/l]
\nonumber\\
&&\qquad\qquad\qquad\mbox{with}\quad
E_2\simeq11.678469-i18.961903[\hbar^2/2ml^2]
\end{eqnarray}
for $V_0=1[\hbar^2/2ml^2]$.
We also use $a_1=a_2=1$ for simplicity.
The fleeing speed~(\ref{eq260}) depends on $x$ as shown in Fig.~\ref{fig5}(a).
The speed is negative near the origin, but is positive all through the region $|x|>l$, converging to $\mathop{\mathrm{Re}}K_2\hbar/m$ fairly quickly.
We show in Fig.~\ref{fig5}(b) the numerical solution $L(t)$ and $\dot{L}(t)$ of the differential equation~(\ref{eq290}) with the initial condition $L(0)=l$.
The solution $L(t)$ behaves almost linearly after some time with $\dot{L}(t)$ converging to $\mathop{\mathrm{Re}}K_2\hbar/m$.
This demonstrates that the resonance with greater $|\mathop{\mathrm{Im}}K_n|$ quickly dominates the behavior.

\section{Summary}
In the present paper, we first reviewed the definition of the resonant and anti-resonant states as eigenstates of the time-independent Schr\"{o}dinger equation.
These states are eigenstates with the Siegert condition, that is, the boundary condition that there are out-going waves only.
We demonstrated for a square-well potential that the resonant and anti-resonant states are indeed obtained under the Siegert condition.

We then presented a physical view of the resonant and anti-resonant states.
In the resonant states, the particle number around the central region exponentially decays in time because of momentum leaks from the region.
The anti-resonant states are the time reversal of the resonant states.
In this sense, the pair of a resonant and anti-resonant states spontaneously breaks the time-reversal symmetry~\cite{Petrosky88,Petrosky97a,Petrosky97b}.

The above physical view led us to the argument of the particle-number conservation for the resonant and anti-resonant states.
In either case, one must consider both spatial and temporal behavior of the resonance wave function in order to have a probabilistic interpretation thereof. 
When we expand the integration region following the momentum leaks, the particle number in the expanding region is conserved.
We then extended the probabilistic interpretation to a superposition of resonances. The latter extension, as well as details of the calculation in Sec.~4 appear here for the first time. Finally, we comment that the next obvious step would be have a clear probabilistic interpretation of superpositions of bound states and resonant states.

The work of NH is supported by Grant-in-Aid for Scientific Research  
No.~17340115 from the Ministry of Education, Culture, Sports, Science  
and Technology as well as by Core Research for Evolutional Science and  
Technology (CREST) of Japan Science and Technology Agency.
The work of JF was supported in part by the Israel Science Foundation (ISF).

\end{document}